\documentclass{jfm}
\usepackage{natbib}
\usepackage{epsfig}
\usepackage{amsmath}
\usepackage{amssymb}
\usepackage{amsfonts}
\usepackage{amsxtra}
\usepackage{color}

\title[Shear-Improved Smagorinsky Model]{Shear-Improved
  Smagorinsky Model for Large-Eddy Simulation of \\ Wall-Bounded
  Turbulent Flows} \author[E. L\'ev\^eque, F. Toschi, L. Shao and
J.-P. Bertoglio]{E. L\'ev\^eque$^1$, F. Toschi$^{2,3}$, L. Shao$^4$
  and J.-P. Bertoglio$^4$}
\affiliation{
  $^1$ Laboratoire de Physique, \textsc{Cnrs}, \'Ecole normale sup\'erieure de Lyon,  France.\\
  $^2$ Istituto per le Applicazioni del Calcolo, \textsc{CNR}, Viale del Policlinico 137, I-00161, Roma, Italy.\\
  $^3$ INFN, Sezione di Ferrara, Via G. Saragat 1, I-44100 Ferrara, Italy.\\
  $^4$ Laboratoire de M\'ecanique des Fluides et d'Acoustique,
  \textsc{Cnrs}, \'Ecole centrale de Lyon, France.  
} 
\date{\today}

\begin{document}
\bibliographystyle{jfm}

\maketitle
\begin{abstract}
  A shear-improved Smagorinsky model is introduced based on recent results concerning shear effects in wall-bounded turbulence by 
  \cite{ShearToschi}. The Smagorinsky eddy-viscosity is modified as
  $\nu_T = (C_s \Delta)^2 (|\overline S|-|\langle \overline
      S\rangle|)$: the magnitude of the mean shear
  $|\langle\overline S\rangle|$ is subtracted to the magnitude of the
  instantaneous resolved strain-rate tensor $|\overline S|$; here $C_S$
  is the standard Smagorinsky constant and $\Delta$ denotes the grid spacing.  This subgrid-scale model is
  tested in large-eddy simulations of plane-channel
  flows at Reynolds numbers $Re_\tau=395$ and $Re_\tau=590$.
  First comparisons with the dynamic Smagorinsky model and direct numerical simulations, including mean velocity, turbulent kinetic energy and Reynolds stress profiles, are shown to be extremely satisfactory.
  The proposed model, in addition of being physically sound, has a
  low computational cost and possesses a high potentiality of
  generalization to more complex non-homogeneous turbulent flows.
\end{abstract}

\maketitle
\section{Introduction}
The prohibitive cost of direct numerical simulations (\textsc{Dns})
of turbulent engineering flows motivate the elaboration of simplified
models, requiring less computation but still relevant (to
some degree) for reproducing the large-scale dynamics
\cite[]{Deardorff,LesieurMetais,LESchallenges,SagautBook}.  In this context, the modeling near a solid boundary is of particular interest.  
A boundary affects the kinetics of the flow through different
mechanisms; the most prominent is that related to the mean shear,
which is extremal in the proximity of the boundary and 
responsible for the production of streamwise vortices and streaky
structures. These fluid structures eventually detach from the boundary and sustain
turbulence in the bulk \cite[]{PerotMoin}.  Thus, it is thought that understanding how the mean shear impacts on fluid motions is a key to improving the capabilities of
numerical models of turbulent flows.  Turbulence near a boundary is not homogeneous nor isotropic and therefore, the customary theoretical background for
modeling the unresolved small-scale dynamics should be surpassed \cite[]{PopeBook}.  In the
present work, we present a new subgrid-scale model of turbulence which may be viewed as
an improvement over the popular Smagorinsky model in the presence of a mean shear.  Our
model originates from recent theoretical findings concerning 
shear effects in wall-bounded turbulence, see
\cite{ShearToschi}. We propose to subtract the magnitude of the
mean shear from the resolved strain-rate tensor in the definition of
the eddy-viscosity. As this will be discussed later, this simple improvement
accounts for the non-isotropic nature of the flow near the boundary and
at the same time, allows us to recover  the standard Smagorinsky model
in regions of (locally) homogeneous and isotropic turbulence.  
The general framework of the so-called large-eddy simulation (\textsc{Les}) of
turbulent flows is now briefly recalled.

Roughly speaking, large-scale motions transport most of the kinetic
energy of the flow. Their strength make them the most efficient
carriers of conserved quantities (momentum, heat, mass, etc.). On the
contrary, small-scale motions are primarily responsible for the
dissipation while they are weaker and contribute little to transport. From
mechanical aspects, the large-scale (energy-carrying) dynamics are
thus of particular importance and the costly computation of small-scale
dynamics should be avoided.  Furthermore, while large-scale motions
are strongly dependent on the external flow conditions, small-scale
motions are expected to behave more universally. Hence, there is a
hope that numerical modeling can be feasible and/or require few
adjustments when applied to various flows. 

In \textsc{Les}, only the large-scale components of flow variables are explicitly integrated in time; interactions with the unresolved small-scale components are
modeled. A spatial filtering is conceptually
introduced as $ \overline{\phi}(\mathbf{x},t)= \int
\phi(\mathbf{x^\prime},t) G_\Delta(\mathbf{x}- \mathbf{x^\prime})
d\mathbf{x^\prime}$, where the filter width $\Delta$ fixes the size of the smallest scales of variation retained in the flow variable ${\phi}(\mathbf{x},t)$
\cite[]{Leonard1974}. In practical applications, $\Delta$ is chosen
much larger than the spatial cutoff scale of $\phi(\mathbf{x},t)$, i.e. the dissipative scale of turbulence, so that
$\overline \phi(\mathbf{x},t) $ may be considered as the large-scale
 component of $\phi(\mathbf{x},t)$.
Applying the previous filtering procedure to the Navier-Stokes equations (and neglecting here non-commutation errors \cite[]{GhosalMoin} for the sake of simplicity) yields
\begin{equation}
  \label{eq:ns_filtered1} 
  \frac{\partial \overline{u}_i}{\partial t} +\overline{u}_j  \frac{\partial \overline{u}_i}{\partial x_j} + \frac{\partial \tau_{ij}}{\partial x_j} 
  = -\frac{\partial \overline{p}} {\partial x_i }+ \nu \frac{\partial^2 \overline{u}_i}{\partial x_k  \partial x_k }\;\;\;\;\mathrm{with} \quad  \frac{\partial \overline{u}_i}{\partial x_i}  = 0 
\end{equation}
where $\overline{u}_i(\mathbf{x},t)$ and $\overline{p}(\mathbf{x},t)$
represent  the large-scale velocity and pressure, and
$\nu$ is the kinematic viscosity of the fluid.
The equations (\ref{eq:ns_filtered1}) are
amenable to numerical discretisation with a grid spacing comparable to
$\Delta$, since $\overline{\mathbf{u}}(\mathbf{x},t)$ is expected to
vary smoothly over the distance $\Delta$.   
$\tau_{ij}(\mathbf{x},t) \equiv \overline{u_i(\mathbf{x},t)
  u_j(\mathbf{x},t)}-\overline{u}_i(\mathbf{x},t)
\overline{u}_j(\mathbf{x},t)$ is named the subgrid-scale
(\textsc{Sgs}) stress tensor and encompasses all interactions between the grid-scale component and the unresolved subgrid-scale component
of ${\mathbf{u}}(\mathbf{x},t)$. In   \textsc{Les},
$\tau_{ij}(\mathbf{x},t)$ needs to be expressed in terms of the
grid-scale velocity field $\overline{\mathbf{u}}(\mathbf{x},t)$ only, which
is the hard problem \cite[]{LesieurBook}.  
Eddy-viscosity
models parameterize the \textsc{Sgs} stress tensor as
\begin{equation}\tau_{ij} - \frac{1}{3} \delta_{ij} \tau_{kk} = -2
  \nu_T \overline{S}_{ij} \;\;\;\;
  {\mbox{where}} \;\;\;\; \overline{{S}}_{ij}(\mathbf{x},t) \equiv \frac{1}{2} \left( \frac{\partial \overline{u_i}}{\partial x_j}(\mathbf{x},t) +
    \frac{\partial \overline{u_j}}{\partial x_i}(\mathbf{x},t) \right)
\label{eq:sgs_stress}
\end{equation} 
where $\nu_T(\mathbf{x},t)$ is the scalar eddy-viscosity and $
\overline{{S}}_{ij}$ is the resolved rate-of-strain tensor.  This
empirical modelization is rooted in the idea that \textsc{Sgs} motions
are primarily responsible for a diffusive transport of momentum from
the rapid to the slow grid-scale flow-regions. 
The theoretical basis for the introduction of an
eddy-viscosity is rather insecure, however, it appears to be workable
in practice (as advocated by \cite{kraichnan_eddy}). The eddy-viscosity $\nu_T$ is then
primarily designed to ensure the correct mean drain of kinetic energy
from the grid-scale flow to the \textsc{Sgs} motions: $-\langle{\tau}_{ij}\overline{{S}}_{ij} \rangle$ 
from the equations (\ref{eq:ns_filtered1}). Another important feature is that $\nu_T(\mathbf{x},t)$ should vanish in laminar flow-regions, e.g. in the viscous sub-layer near the boundary \cite[]{MoinKim1982}. Accordingly, our main concern was to determine an eddy-viscosity
$\nu_T(\mathbf{x},t)$ that would take into account shear effects in
the transfer of kinetic energy to the \textsc{Sgs} motions (without any sort of dynamical adjustment)
and naturally decrease to zero at the boundary (without using any ad-hoc damping function).

In Section \ref{sec:refined_smagorinsky}, the advantages and drawbacks
of the Smagorinsky model, which is used as our baseline model, are
recalled.  The necessity for a new model that can achieve a satisfactory
compromise between accuracy and manageability is identified. A
shear-improved Smagorinsky model is introduced within the classical
picture of shear turbulence.  In Section \ref{sec:LES}, we present
results from \textsc{Les} of turbulent
plane-channel flows at $Re_\tau=395$ and
$Re_\tau=590$. Comparisons are carried out with the dynamic
Smagorinsky model and \textsc{Dns}. Discussion and
perspectives follow in Section \ref{sec:conclusion}.


\section{A shear-improved Smagorinsky model \label{sec:refined_smagorinsky}}
\subsection{Our baseline model: The Smagorinsky
  model \label{ssec:smagorinsky}}
  
The Smagorinsky model is certainly the simplest and most commonly used
eddy-viscosity model (see \cite{smagorinsky}, \cite{PopeBook} for a comprehensive description). The prescription for $\nu_T$ writes 
$\nu_T(\mathbf{x},t) = (C_s \Delta)^2 
|\overline{S}(\mathbf{x},t)|$, where $|\overline{S}| \equiv (2
  {\overline{S}_{ij}}{\overline{S}_{ij}})^{1/2}$ represents the magnitude of the
resolved rate-of-strain and $C_s$ is a non-dimensional coefficient (called the
Smagorinsky constant).  The major merits of the Smagorinsky model are
its manageability, its computational stability and the simplicity of
its formulation (involving only one adjusted parameter). All this makes it
a very valuable tool for engineering applications \cite[]{RogalloMoin}.  However, while
this model is found to give acceptable results in the \textsc{Les} of
homogeneous and isotropic turbulence (with $C_s \approx 0.17$ according to \cite[]{Lilly}), 
it is found to be too dissipative with respect to the resolved motions in near-wall turbulence, 
due to an excessive eddy-viscosity arising from the mean shear \cite[]{MoinKim1982}. The eddy-viscosity predicted by Smagorinsky is non-zero in laminar flow-regions; the model 
introduces spurious dissipation which has the effect of damping the
growth of small perturbations and thus restrain the transition to
turbulence \cite[]{PiomelliZang}. 

To alleviate these deficiencies in the case of wall-bounded flows, 
the Smagorinsky constant $C_s$ is often multiplied by
a damping factor depending on the wall-normal distance,  
the van Driest function being the prime example \cite[]{vandriest}.  
Although the
van Driest damping function is commonly employed, its theoretical
basis has never been adequately addressed, thereby leaving this
function rather arbitrary.  Moreover, if the determination of the distance is
straightforward in the case of a plane boundary,  it becomes
more ambiguous near a curved boundary or a sharp corner.  A more general
approach, free from the use of the wall-normal distance, is therefore
desirable. 
The dynamic \textsc{Sgs} model intends 
to evaluate the Smagorinsky constant (from
the resolved motions) as the calculation progresses \cite[]{dynamic_model} and thus avoids the need to specify  a priori, or
tune, its value. In brief, the adjustment of $C_s$ relies on the Germano identity \cite[]{Germano}
and assumes the scale similarity of resolved velocity fluctuations at scales comparable to $\Delta$
\cite[]{review_les}. This methodology yields a coefficient $C_s$ that varies with
position and time and vanishes in the
vicinity of a solid boundary with the correct behavior \cite[]{Piomelli_data}. 
It is beyond dispute that
the dynamic procedure greatly improves the capacity of the
original Smagorinsky model, however, this progress
is also accompanied with a certain deterioration of the numerical stability and an increase of the computational cost. 
In this situation, we believe that it is legitimate to seek for a \textsc{Sgs} model that would achieve a better
compromise between accuracy  and manageability: ideally, as simple as
the original Smagorinsky model and as accurate as the dynamic Smagorinsky model. 
The present \textsc{Sgs} model that we shall call \emph{shear-improved Smagorinsky model} 
has been elaborated along this line of idea.

\subsection{Our proposal: A shear-improved
  eddy-viscosity \label{ssec:smagorinsky_improved}}
Our \textsc{Sgs} model differs from the standard Smagorinsky model in
how the eddy-viscosity is defined. It borrows ideas originally advanced by
\cite{schumann1975} (see also \cite{Sullivan}) by using a two-part eddy-viscosity accounting for
the interplay between the non-linear
energy cascade present in isotropic turbulence, and mean shear effects associated with anisotropy.
However, it differs from Schumann's proposal in its formulation, in particular for an extra simplicity. 
Our proposal for the \textsc{Sgs} viscosity writes
\begin{equation}
  \nu_T=(C_s \Delta)^2 \cdot \left( |\overline S|-|\langle\overline S\rangle|\right)
\label{eq:eddy_viscosity}
\end{equation}
where the brackets $\langle~\rangle$ would a priori denote an ensemble average, in practice,
space average over homogeneous directions and/or time average will be used (this specific issue will be mentioned again in Section \ref{sec:conclusion}). From the definition (\ref{eq:eddy_viscosity})  
the mean drain of kinetic energy from the grid-scale to the subgrid-scale motions is given by
\begin{equation}{F_\mathrm{sgs}} \equiv -\langle\tau_{ij} \overline S_{ij} \rangle = (C_s
\Delta)^2 ~ \left( \langle |\overline S|^3\rangle - |\langle\overline
    S\rangle| \langle|\overline S|^2 \rangle \right).
    \label{eq:sgs_flux}
    \end{equation}  
Straightforwardly, our eddy-viscosity vanishes if the
resolved turbulence disappears, i.e. if $\overline  S = \langle\overline
  S\rangle$. We shall argue that it is also consistent, to some extent,
  with the \textsc{Sgs} energy budget of shear turbulence.  The theoretical basis of our model was first put forward by \cite{ShearToschi}, on account of previous numerical and experimental studies on wall-bounded turbulence \cite[]{ToschiPrl1,ToschiPof1,Ruiz}. For the sake of simplicity, we shall here recast the key arguments in the ideal case of a statistically stationary homogeneous shear flow \cite[]{moninyaglom}.

In an homogeneous shear flow, the velocity field $\mathbf{u}(\mathbf{x},t)$
may be decomposed into $u_i(\mathbf{x},t) = u_i^\prime(\mathbf{x},t) + ({\partial
  U_i}/{\partial x_j})x_j,$ where $U_i(\mathbf{x})$  and $u_i^\prime(\mathbf{x},t)$ denote the
mean  and  fluctuating part of the velocity, respectively.  Starting from the exact dynamical equations for the two-point correlation function $R(r)=\langle u_i^\prime(\mathbf{ x},t)
  u_i^\prime(\mathbf{ x}+\mathbf{ r},t) \rangle$ and by integrating this equation over
a sphere $B_r$ of radius $r$ centered at $\mathbf{x}$, one can
establish an energy budget (at scale $r$) which takes the form (see \cite{Casciola,danaila2004}):
\begin{equation}
S_3^{\mathrm{tr}}(r) + S_3^{\mathrm{pr}}(r) = - \frac 4 3 \varepsilon r + 2 \nu \frac {\mathrm{d}} {\mathrm{d}r} \left( \frac 1 {4 \pi r^2 }\oint_{\partial B_r} \langle  |\delta \mathbf{u}^\prime(\mathbf{x},\mathbf{r},t)|^2 \rangle ~\mathrm{dS} \right).
\label{eq:budget}
\end{equation}
 
The two contributions $S_3^{\mathrm{tr}}(r)$ and $S_3^{\mathrm{pr}}(r)$ arise
from the non-linear term of the Navier-Stokes equations. In the right-hand side of (\ref{eq:budget}) 
 $\varepsilon$ denotes the mean energy dissipation rate and the second term encompasses finite Reynolds
effects (at scale $r$) with  $\delta \mathbf{u}^\prime(\mathbf{x},\mathbf{r},t) \equiv \mathbf{u}^\prime(\mathbf{x}+\mathbf{ r},t)-\mathbf{u}^\prime(\mathbf{x},t)$.  The energy budget (\ref{eq:budget}) is the generalization of the Karman-Howarth equation in the context of homogeneous shear turbulence
\cite[]{HinzeBook}. In the context of \textsc{Les}, it may also be
interpreted as the proper mean \textsc{Sgs} energy budget with respect
to the grid scale $r$.  More explicitly, \begin{equation}
S_3^{\mathrm{tr}}(r) = \frac 1{4 \pi r^2} \oint_{\partial B_r} \left(
  \langle |\delta \mathbf{u}^\prime(\mathbf{x},\mathbf{r},t)|^2 \delta u_i^\prime(\mathbf{x},\mathbf{r},t) \rangle
  +  \frac {\partial U_i}{\partial x_j} r_j\langle |\delta \mathbf{u}^\prime(\mathbf{x},\mathbf{r},t)|^2 \rangle
  \right) \mathrm{dS}_i
\label{eq:transfer}
\end{equation} represents the transfer of kinetic energy from grid-scale
motions (at scales larger than $r$) to subgrid-scale motions.  Also,
it indicates that this transfer results from both the grid-scale
turbulent fluctuations and the mean shear. 
These two effects correspond respectively to the non-linear triple-correlation term and the rapid (or linear) term entering in the spectral decomposition of the energy transfer, as first evidenced by \cite{Craya}.
 The
second term in the left-hand side of (\ref{eq:budget}) represents a
production of \textsc{Sgs} kinetic energy induced by the 
mean shear, and is expressed as \begin{equation}S_3^{\mathrm{pr}}(r) = \frac 1 {4 \pi r^2}
\int_{B_r} 2 \frac {\partial U_i}{\partial x_j} \langle \delta
  u^\prime_i(\mathbf{x},\mathbf{r},t) \delta u^\prime_j(\mathbf{x},\mathbf{r},t) \rangle \mathrm{dV}
\label{eq:production}.\end{equation}  In summary, the energy budget
(\ref{eq:budget}) means that in the steady-state, the \textsc{Sgs}
dynamics  are sustained against molecular dissipation 
(represented by the right-hand side of (\ref{eq:budget})) by the transfer
of energy originated from grid-scale motions ($S_3^{\mathrm{tr}}(r)$) and the production of
energy induced directly by the mean shear ($S_3^{\mathrm{pr}}(r)$). 

As mentioned in the introduction, the estimate of the \textsc{Sgs}
energy flux, which should be identified with
$S_3^{\mathrm{tr}}(\Delta)/\Delta$ in the previous developments, is of prime importance in the modeling of the eddy-viscosity. 
Roughly speaking, one may consider that turbulent grid-scale velocity differences $\delta
u^\prime(\Delta)$ typically behave as the fluctuating part of the
resolved rate-of-strain $|\overline{S^\prime}|$ multiplied by $\Delta$, i.e.  $\delta
u^\prime(\Delta) \approx |\overline{S^\prime}| \Delta$. In the same way, $\delta U(\Delta) \approx |\langle \overline{S} \rangle|\Delta$.
Equation (\ref{eq:transfer}) then suggests that the mean \textsc{Sgs} energy flux should involve 
two separate contributions of order $\Delta^2 \langle |\overline {S^\prime}|^3\rangle$
and $\Delta^2 | \langle \overline S \rangle| \langle | \overline
  {S^\prime} |^2\rangle$, respectively.  
  
  In flow-regions where $|\overline{S^\prime}| \gg |\langle \overline S
\rangle|$, the \textsc{Sgs} energy flux reduces to the contribution of order
$\Delta^2 \langle |\overline {S^\prime}|^3\rangle$. In these regions, the mean shear is too weak to perturb the grid-scale dynamics; eddies of size comparable to the grid-scale $\Delta$ adjust dynamically via non-linear interactions to transfer energy to \textsc{Sgs} motions. This is the standard mechanism behind homogeneous and isotropic turbulence \cite[]{frisch}.
From our expression (\ref{eq:sgs_flux}), we remark that the Smagorinsky estimate   
${F_\mathrm{sgs}} \simeq (C_s \Delta)^2  \langle |\overline{S^\prime} |^3\rangle $ is consistently recovered in that case. 
In regions where $|\langle \overline S \rangle| \gg |\overline{S^\prime}|$, the behavior of the flow is clearly different:  
eddies of size comparable to the grid-scale have no
time to adjust dynamically and are rapidly distorted by the mean shear \cite[]{LiuKatz}. In these regions, the \textsc{Sgs} energy flux is driven by the mean shear and therefore is dominated by the contribution of order $\Delta^2 | \langle \overline S \rangle| \langle | \overline  {S^\prime} |^2\rangle$. From expression (\ref{eq:sgs_flux}) and  assuming that 
$\langle |\overline{S} |^3\rangle \approx \langle |\overline{S} |^2\rangle^{3/2}$, one obtains 
${F_\mathrm{sgs}} \simeq 1/2 ~(C_s \Delta)^2  |\langle \overline S \rangle|~\langle |\overline{S^\prime}|^2\rangle$ in agreement with the previous reasoning. We may thus conclude that proposal (\ref{eq:eddy_viscosity}) for the eddy-viscosity is consistent with the \textsc{Sgs} energy budget of (locally homogeneous) shear turbulence in the two limiting situations $|\overline{S^\prime}| \gg |\langle \overline S \rangle|$ and $|\langle \overline S \rangle| \gg |\overline{S^\prime}|$. The first results concerning plane-channel flows (see Section \ref{sec:LES}) indicate that the proposed model abridges between these two situations without the need for any additional adjustment.

\section{\textsc{Les} of turbulent plane-channel flows\label{sec:LES}}
Over the last twenty years \textsc{Les} of wall-bounded flows have
received a considerable attention (see \cite{PiomelliBalaras} for a recent review), with the turbulent plane-channel flow (\cite{channel_flow_kim} for an original work) being  the prototypical case. This flow allows for the investigation of shear effects in a simple geometry and has therefore provided a useful test bed of our eddy-viscosity model.  Furthermore,
we have been able to confront our results, including mean velocity, turbulence intensities and Reynolds stress profiles, with the well-established 
literature present on that case, e.g. the comprehensive
\textsc{Dns} database obtained by \cite{moserdb} or, more recently, by \cite{hoyasjimenez}.

\subsection{Numerical simulations}

We performed two \textsc{Les} at $Re_\tau=395$ and $Re_\tau=
590$, where $Re_\tau$ is the Reynolds number based on the friction
velocity $u_\tau$: $Re_\tau \equiv {u_\tau H}/{\nu}$ ($H$ is the half
width of the channel). These two cases correspond to the \textsc{Dns}
conducted by \cite{moserdb}. Periodic
boundary conditions were imposed in the streamwise and spanwise
directions; no-slip conditions at the wall.  The flow was
simulated by integrating the filtered Navier-Stokes equations
(\ref{eq:ns_filtered1}) with the prescribed eddy-viscosity (\ref{eq:eddy_viscosity}).  The right-hand side
of the equations (\ref{eq:ns_filtered1}) was supplemented by an (external)
pressure-gradient ${\Delta {p}_{\textrm{\scriptsize ext}}}/
{4\pi H}$ in order to drive the flow in the streamwise direction. The
pressure difference $\Delta {p}_{\textrm{\scriptsize ext}}(t)$ was
adjusted dynamically to keep a constant flow rate through the channel.
The integration in time relied on a Fourier-Chebyshev pseudo-spectral solver
(de-aliased by using the $3/2$ rule) based on a third-order Runge-Kutta scheme. 
More details about the numerical simulations can be found in \cite{details_simulation}.
The Smagorinsky constant was fixed to its standard value $C_s=0.16$ (in the case of homogeneous and isotropic turbulence) and the scale $\Delta$ was estimated as $(\Delta x\Delta y\Delta z)^{1/3}$, where $\Delta x$, $\Delta y$ and $\Delta z$ denote the local grid sizes in each direction. In the following, 
the grid-scale velocity components are  $U+u^\prime$, $v^\prime$ and $w^\prime$ along the streamwise, wall-normal and spanwise directions, respectively. 

\subsection{Numerical results}

\begin{figure}
\begin{center}
  \epsfig{file=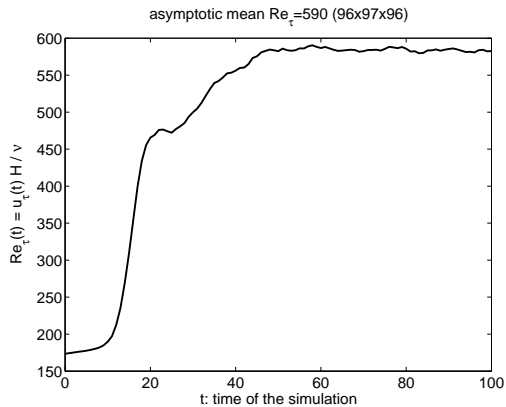,width=6.7cm}
  \caption{The time development of the Reynolds number $Re_\tau(t)$ based on the plane-averaged friction velocity $u_\tau(t)$. }
  \label{fig:transition}
\end{center}
\end{figure}

At initial time, velocity distributions were designed to satisfy the Poiseuille profile plus a small random perturbation. The time development of  $Re_\tau(t)=u_\tau(t)H/ \nu$, where $u_\tau(t)$ expresses as the square root of the plane-averaged wall shear stress, is plotted in Figure \ref{fig:transition} for the case $Re_\tau=590$. A transition (or drag crisis) to the appropriate turbulent regime occurs naturally as the integration (involving the eddy-viscosity (\ref{eq:eddy_viscosity})) progresses. This feature constitutes a first improvement over the Smagorinsky model, for which such transition is not captured. Once a developed turbulent state was achieved, statistics were accumulated.

\begin{figure}
\begin{center}
  \epsfig{file=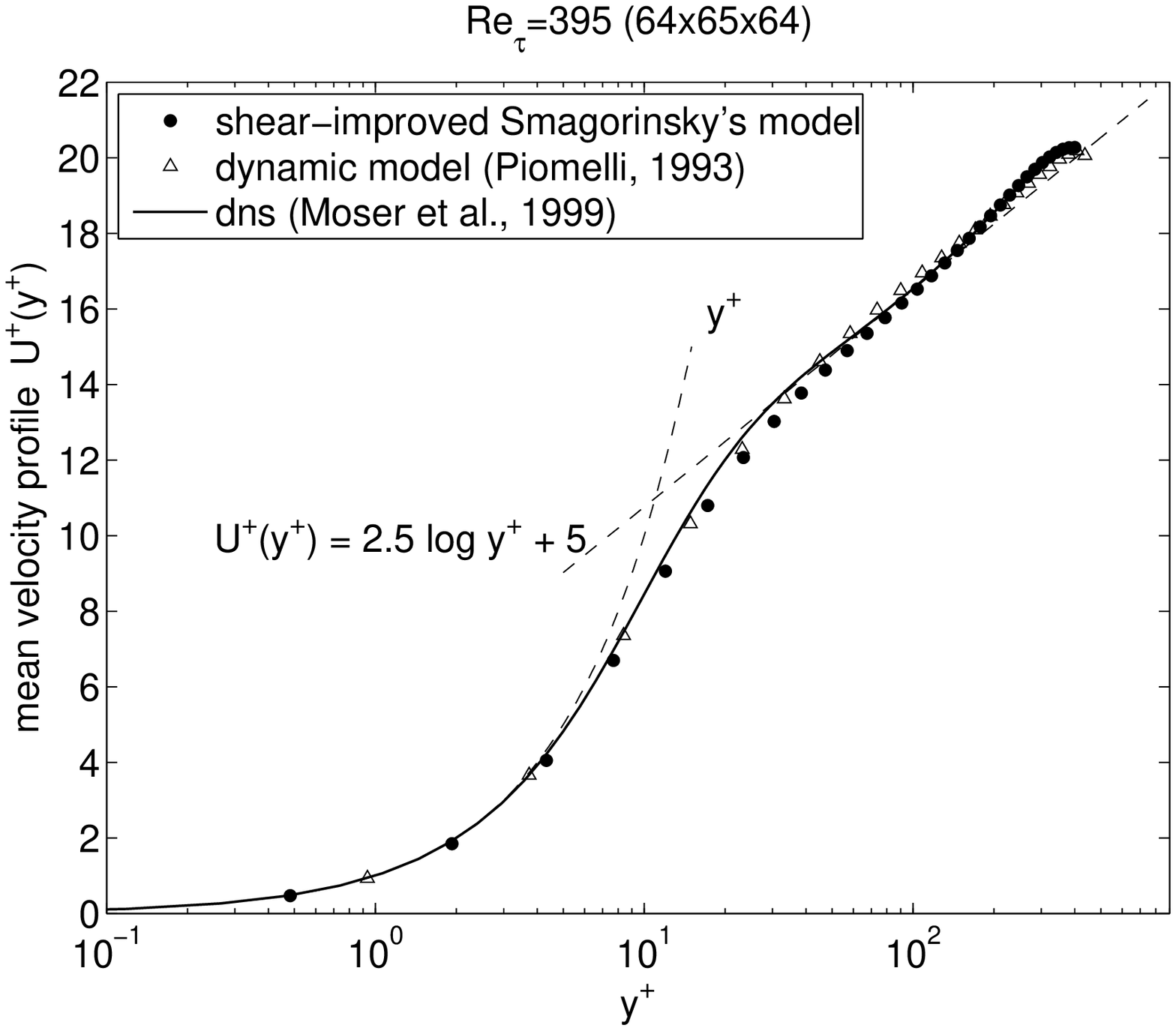,width=6.7cm}
  \epsfig{file=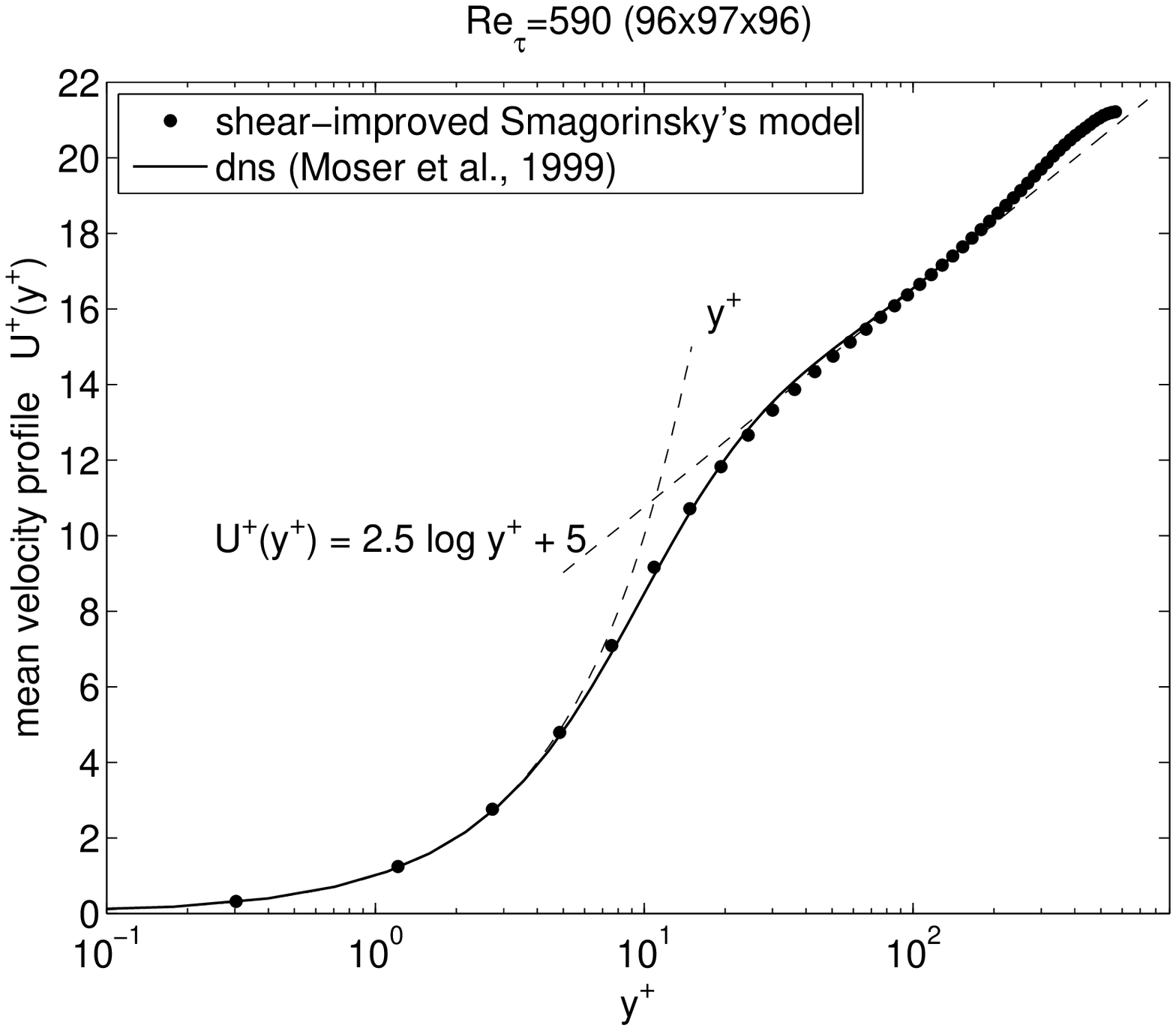,width=6.7cm}
  \caption{Left: ($\bullet$) mean-velocity profile (in wall units) at
    $Re_\tau=395$. The computational domain (in outer units) is $4 \pi
    H \times 2H \times 2 \pi H$ with $64\times 65\times 64$ grid
    points. In comparison with ($-$) the \textsc{Dns} data obtained
    by \protect\cite{moserdb} in the domain $2 \pi H\times 2H\times \pi
    H$ with $256\times 193\times 192$ grid points, and ($\triangle$) a
    computation of the dynamic Smagorinsky model carried out
    by Piomelli (private communication) in the domain $5\pi H/2 \times 2 H
    \times \pi H /2$ with $48 \times 49\times 48$ grid points (using a
    pseudo-spectral solver).  Right: ($\bullet$) mean-velocity profile
    at $Re_\tau=590$ with $96\times 97\times 96$ grid points.
    In comparisons with ($-$) the \textsc{Dns} data with
    $384\times 257\times 384$ grid points. }
  \label{fig:mean}
\end{center}
\end{figure}
 
\begin{figure}
\begin{center}
  \epsfig{file=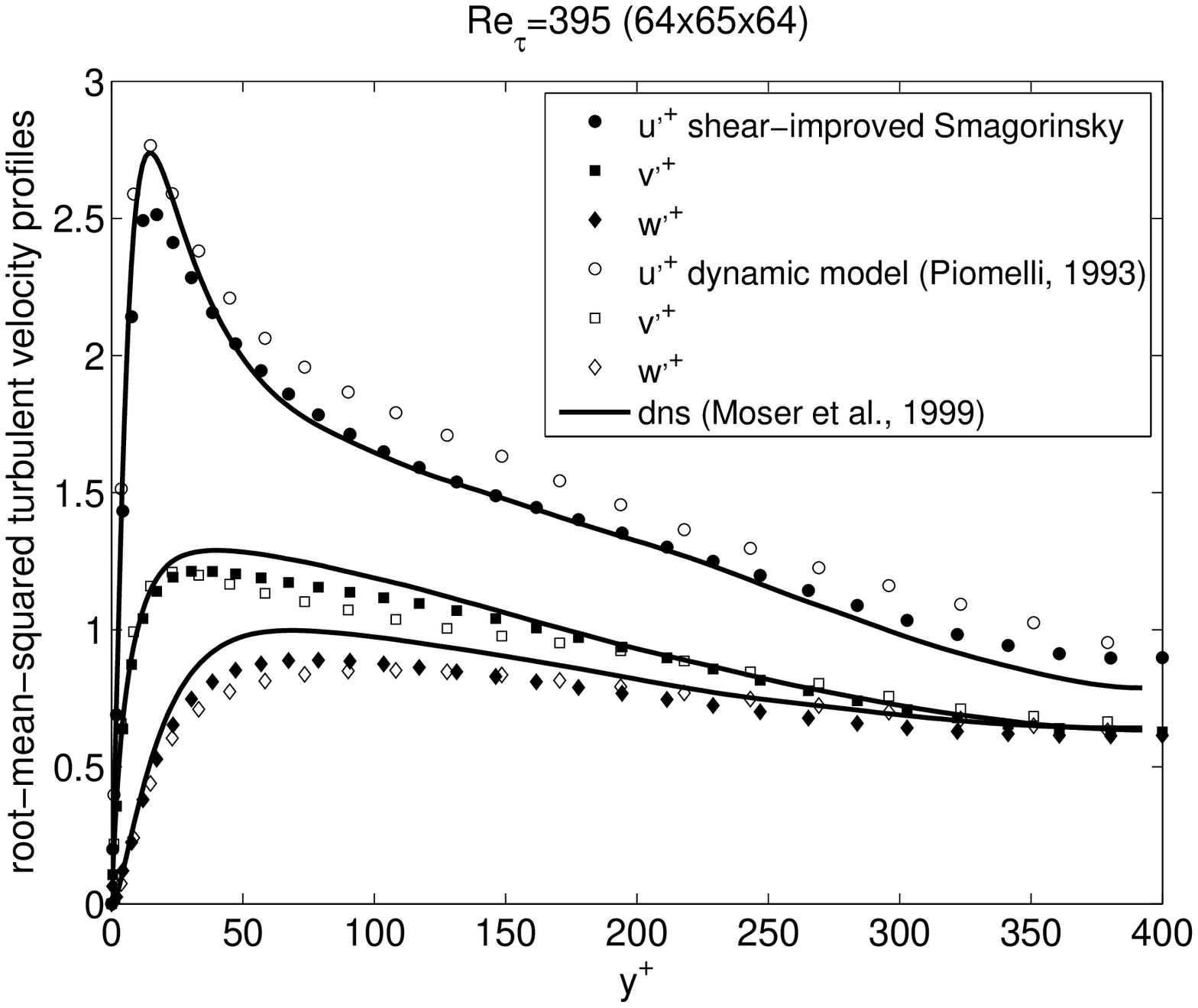,width=6.7cm}
  \epsfig{file=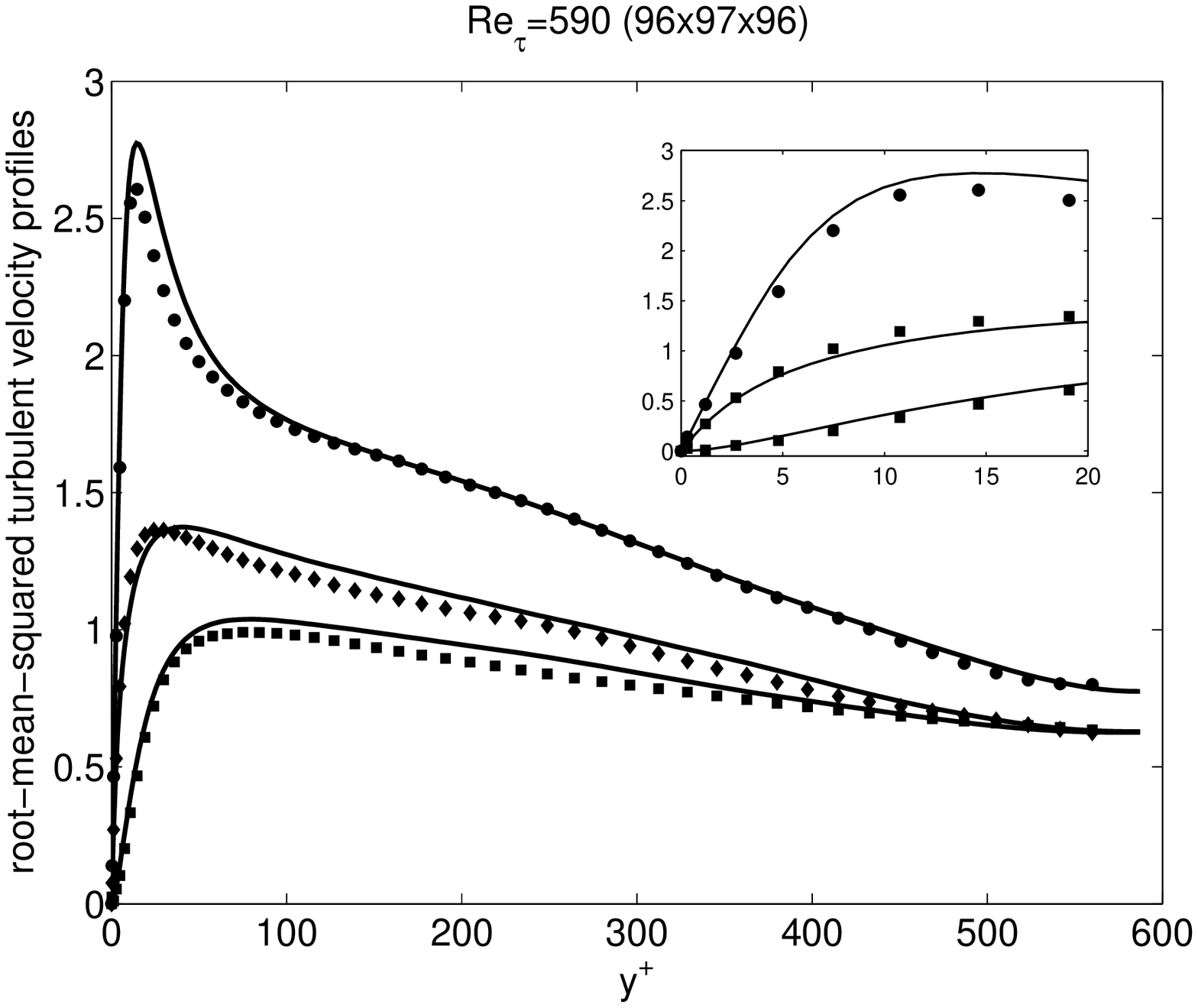,width=6.7cm}
  \caption{Left: Turbulent intensity profiles at $Re_\tau=395$ in comparison with \textsc{Dns} data ($-$) and \textsc{Les} data obtained with the dynamic Smagorinsky \textsc{Sgs} model (open symbols). Right:
    Turbulent intensity profiles at $Re_\tau=590$. The inset focuses on the near-wall behavior in comparison with \textsc{Dns} data.}
  \label{fig:u2v2w2}
\end{center}
\end{figure}

The mean velocity $U^+(y^+)$ is displayed as a function
of the wall-normal distance  $y^+$ in Figure \ref{fig:mean}. From now on, the average is meant
in time and over horizontal $x$-$z$ planes (homogeneous directions). Mean velocity profiles agree very well with the \textsc{Dns} data of
reference obtained by \cite{moserdb} for both considered
 Reynolds numbers. The comparison with the dynamic Smagorinsky model is also very satisfactory \cite[]{Piomelli_data}. 
Turbulence intensity profiles (normalized by the squared friction velocity) are displayed in Figure \ref{fig:u2v2w2}. Also here, the profiles compare very well with the \textsc{Dns} data and \textsc{les} results based on the dynamic Smagorinsky model: notice how peak values are suitably captured. As for the dynamic Smagorinsky model, a slight drop is observed for the spanwise and wall-normal components, in the log-layer; on the contrary, the streamwise component fits remarkably the \textsc{Dns} data. In the proximity of the wall, the correct behavior is obtained. 

\begin{figure}
\begin{center}
  \epsfig{file=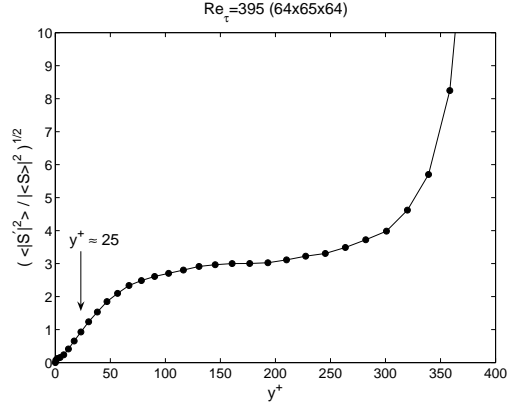,width=6.7cm}
  \caption{The ratio $\sqrt{\langle |   {S^\prime} |^2\rangle } ~/~  |\langle S \rangle|$ is displayed as a function of the wall-normal distance $y^+$ in the \textsc{Les} at $Re_\tau=395$.  For $y^+ \lesssim 25$, the mean shear dominates over the fluctuating part of rate-of-strain. In that region, our eddy-viscosity differs from the original Smagorinsky model candidate. }
  \label{fig:shear}
\end{center}
\end{figure}

These positive results make us presume that our eddy-viscosity has captured the essential of shear effects in wall-bounded turbulence. Now, in order to better check this hypothesis, the ratio
$\sqrt{\langle |   {S^\prime} |^2\rangle } ~/~  |\langle S \rangle|$ is shown as a function of the wall-normal distance in Figure \ref{fig:shear}. For  $y^+ \lesssim 25$, one can see that the mean shear dominates over the fluctuating part of the rate-of-strain, indicating the predominance of the mean-shear component of the eddy-viscosity in that region. Note that the transition distance $y^+ \simeq 25$ is fully consistent with the empirical characteristic distance $A^+=25$ commonly used in the van Driest damping function \cite[]{PopeBook}. In the log-layer, $\sqrt{\langle |   {S^\prime} |^2\rangle } ~/~  |\langle S \rangle|$ increases slowly with $y^+$ (the standard description of the log-layer predicts a linear increase, according to $| \langle {S}\rangle | \propto 1/y$ and $\langle |   {S^\prime} |^2\rangle \propto u_\tau^2/\Delta^2$ \cite[]{SchlichtingBook}) and eventually diverges around the centerline of the channel. Thus, we may claim that our model suitably abridges between the situation where the mean shear prevails (close to the boundaries) and the situation where the fluctuating part of the rate-of-strain dominates (in the bulk of the channel). 

\begin{figure}
  \begin{center} 
    \epsfig{file=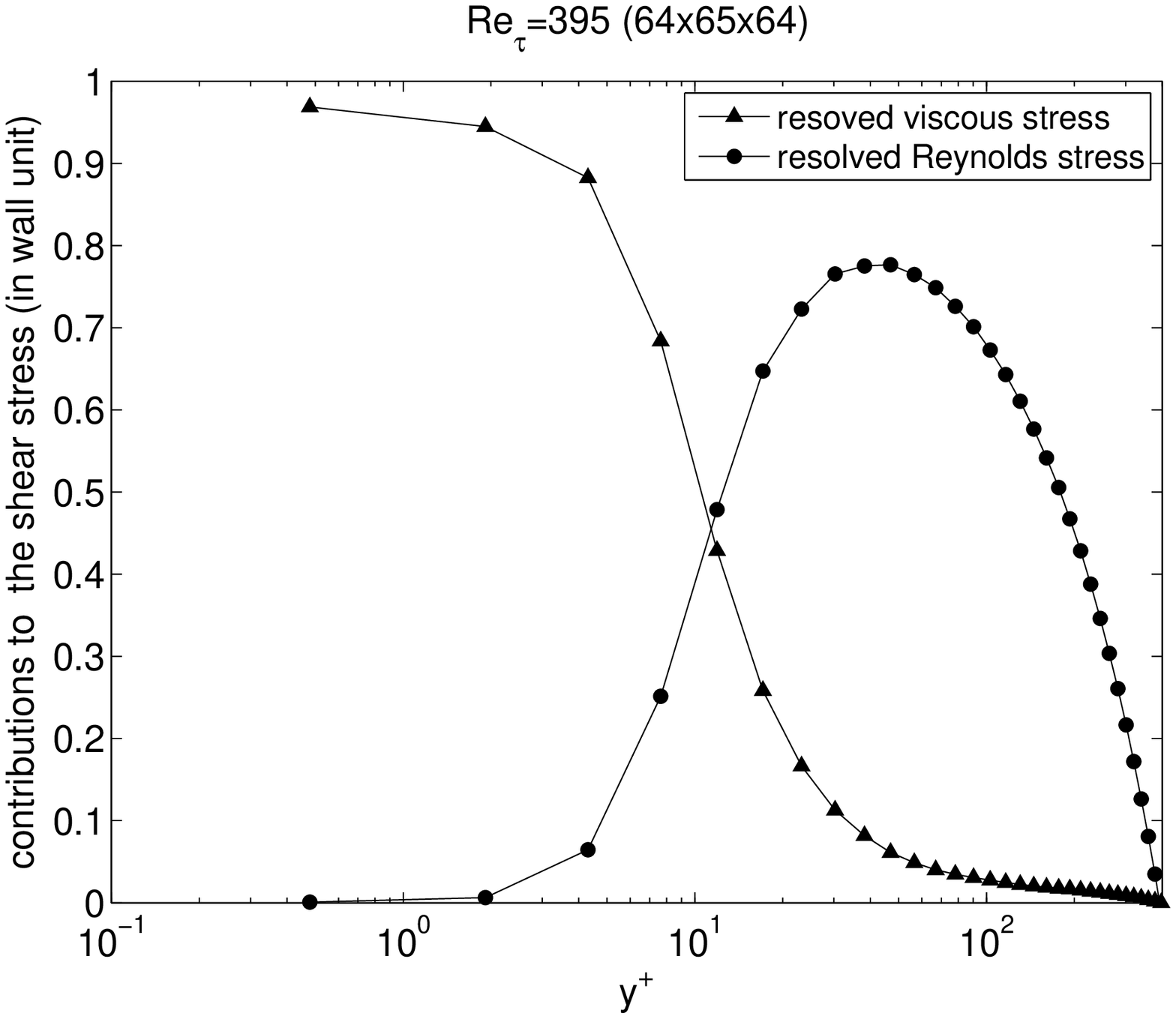,height=5.5cm}
    \epsfig{file=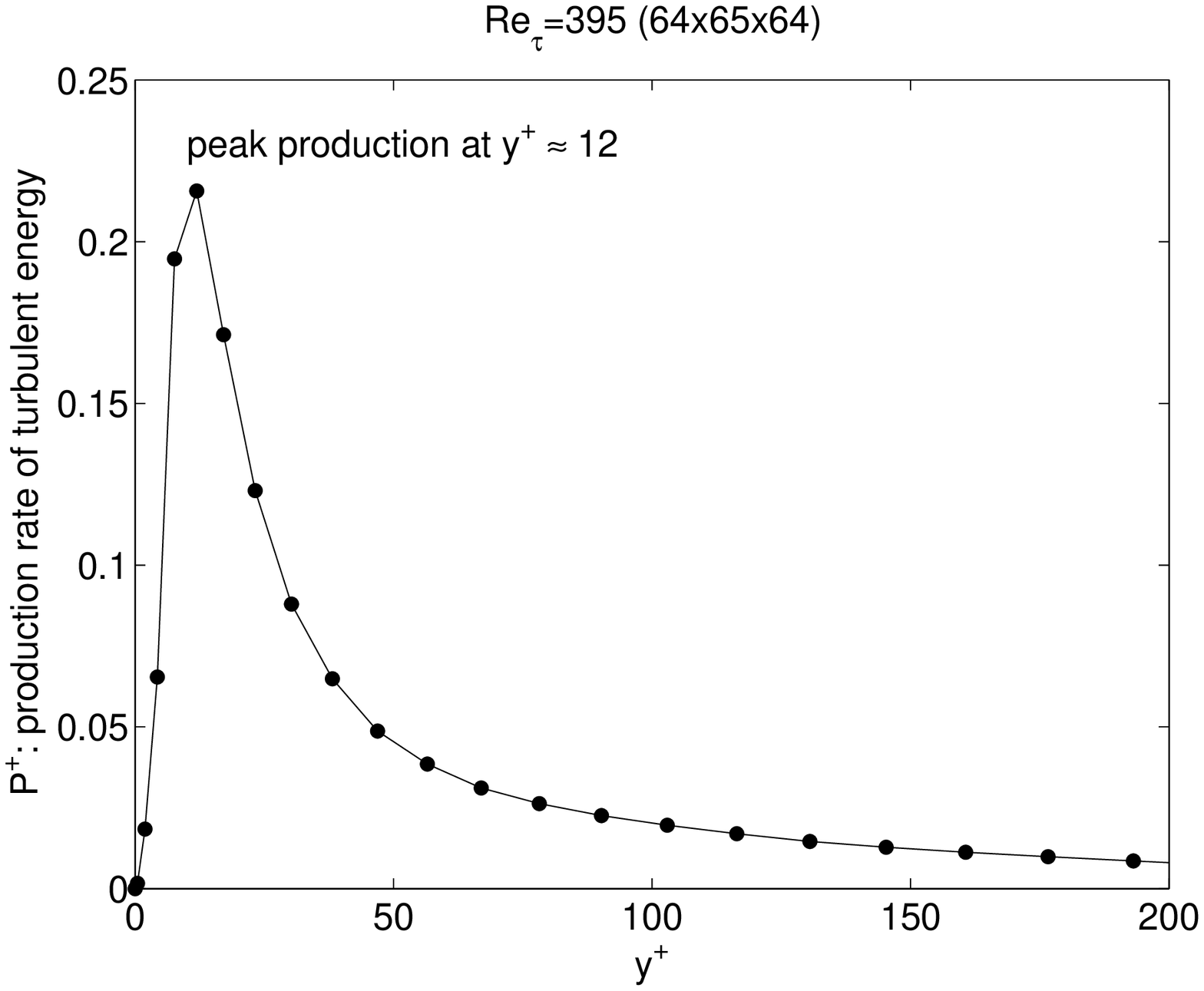,height=5.5cm}
    \caption{Left: The viscous and Reynolds stresses (computed from the
      resolved velocity) at $Re_\tau=395$. The two
      contributions are equal at $y^+
      \approx 12$, in full agreement with \textsc{Dns} results. Right: The peak production of turbulent energy also occurs at $y^+
      \approx 12$, as expected.}
    \label{fig:stress}
\end{center}
\end{figure}

\begin{figure}
  \begin{center} 
    \epsfig{file=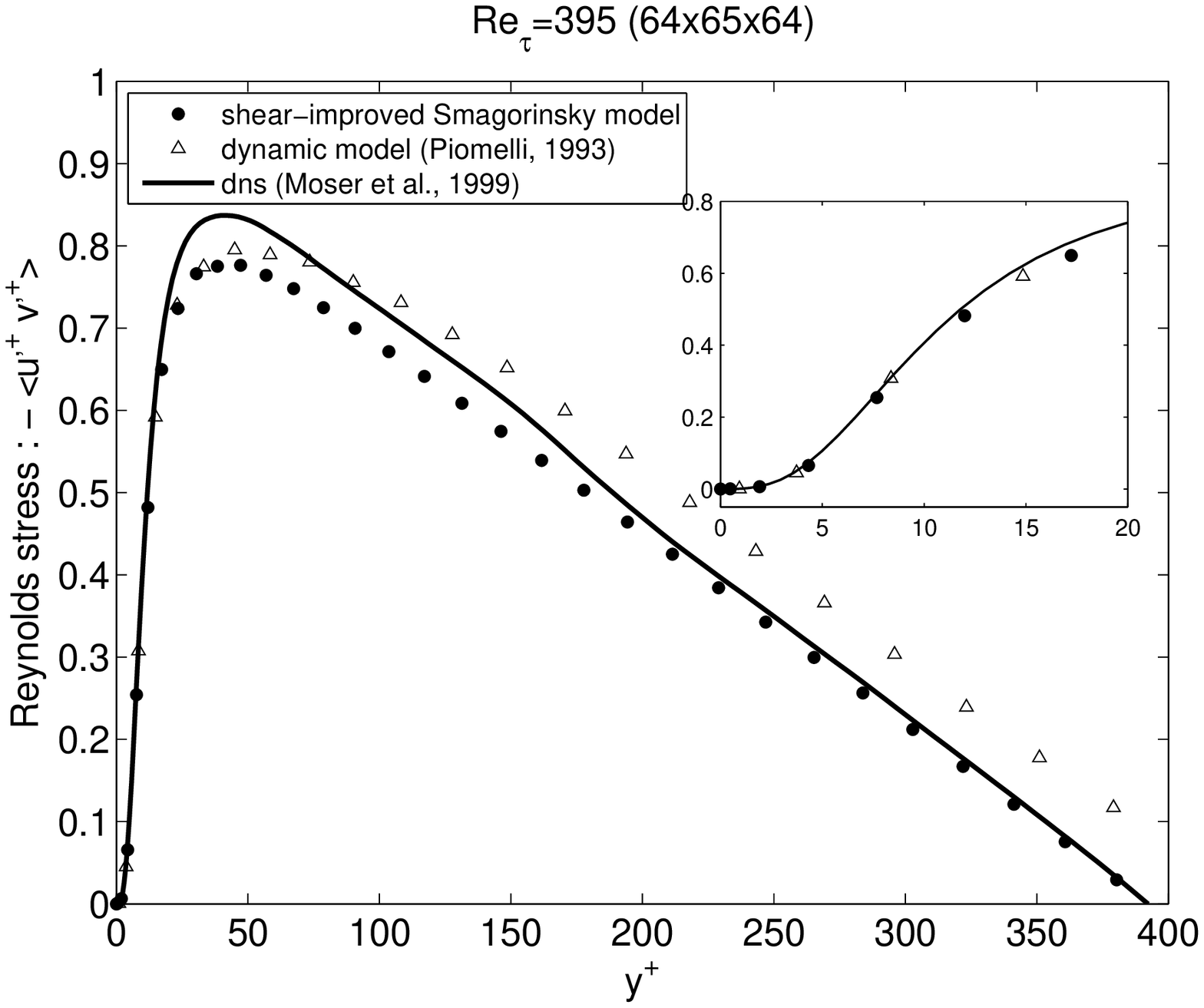,width=6.7cm}
    \epsfig{file=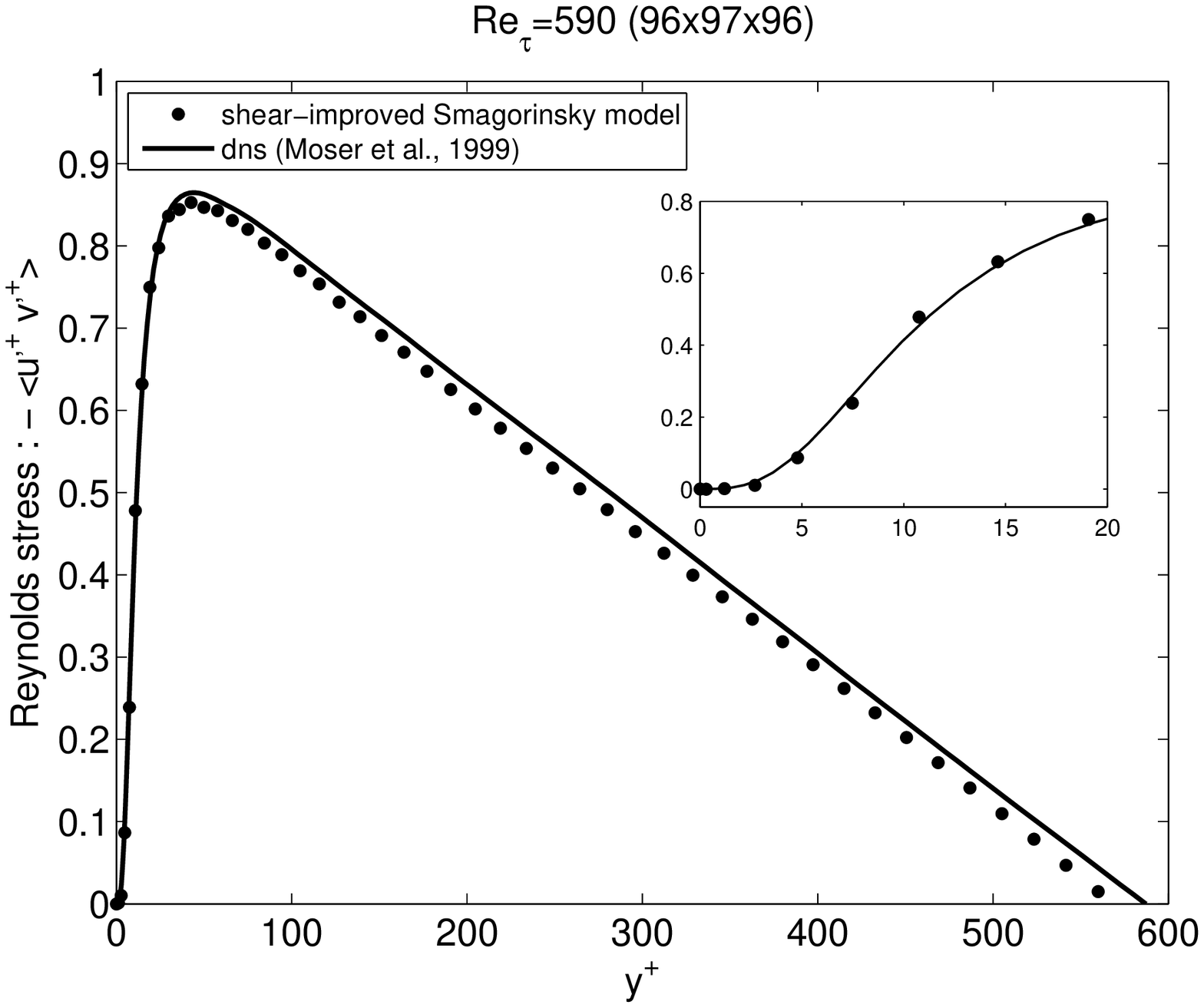,width=6.7cm}
    \caption{Left: Reynolds stress at $Re_\tau=395$ (computed from the resolved velocity). Right: The Reynolds  stress at $Re_\tau=590$. The insets focus on the near-wall behavior.}
    \label{fig:reynolds_stress}
  \end{center}
\end{figure}

The viscous, $dU^+/dy^+$,  and the Reynolds, $\langle u^+v^+ \rangle$, contributions to the resolved stress  are shown in
Figure \ref{fig:stress}. They equal for 
 $y^+ \approx 12$ in perfect agreement with \textsc{Dns} results \cite[]{PopeBook}.
 The peak value of the turbulent energy production, $\langle u^+v^+ \rangle dU^+/dy^+$, occurs at
 the same distance $y^+ \approx 12$, as expected theoretically from the Navier-Stokes equations. 
The grid-scale Reynolds stress profile is displayed in Figure \ref{fig:reynolds_stress}. The agreement with the \textsc{Dns} and the \textsc{Les} based on the dynamic Smagorinsky model is fair and also
the behavior close to the wall is very satisfactory.

\section{Discussions and perspectives\label{sec:conclusion}}

Concluding, we presented a very simple \textsc{Sgs} model consisting in a physically sound improvement over the well-established Smagorinsky model. An explicit connection with the scale-by-scale energy budget of homogeneous shear turbulence has also been identified. Our first results for turbulent plane-channel flows indicate that the proposed model possesses a very good predictive capability (essentially equivalent to the dynamic Smagorinsky model) with a computational cost and a manageability comparable to the original Smagorinsky model.

The generalization to more complex non-homogeneous flows is a priori straightforward since no geometrical argument enters in the definition of the eddy-viscosity. However, an appropriate average  must be specified in the absence of homogeneity directions. As a natural candidate, we suggest an average in time to evaluate the mean components of the rate-of-strain tensor (assuming that the flow is statistically stationary). In the case of non-stationary flows, an ensemble average 
(over several realizations of the flow) may be envisaged. These points are the subject of current investigations in progress.

\acknowledgements{The authors are grateful to Ugo Piomelli for
  providing numerical data; in particular F.T. wishes to thank him for the many useful discussions. Fang Le helped us to perform the 
  \textsc{Les} at the Tsinghua University in China. This work was jointly
  supported by the \'Ecole normale sup\'erieure de Lyon and the
  \'Ecole centrale de Lyon under a \emph{Bonus-Qualit\'e-Recherche}
  grant. Finally, F.T. acknowledges E.L. for his kind hospitality at the \textsc{Ens-Lyon}.  }
\bibliography{papierLESjfm}
\end{document}